# Is dark matter an illusion created by the gravitational polarization of the quantum vacuum?

Dragan Slavkov Hajdukovic[1]
PH Division CERN
CH-1211 Geneva 23
dragan.hajdukovic@cern.ch
[1]On leave from Cetinje, Montenegro

**Abstract**

Assuming that a particle and its antiparticle have the gravitational charge of the opposite sign, the physical vacuum may be considered as a fluid of virtual gravitational dipoles. Following this hypothesis, we present the first indications that dark matter may not exist and that the phenomena for which it was invoked might be explained by the gravitational polarization of the quantum vacuum by the known baryonic matter.

Let us start with a major unresolved problem. The measured galaxy rotation curves remain roughly constant at large radii. Faster than expected orbits, require a larger central force, which, in the framework of our theory of gravity, cannot be explained by the existing baryonic matter. The analogous problem persists also at the scale of clusters of galaxies.

The favoured solution is to assume that our current theory of gravity is correct, but every galaxy resides in a halo of dark matter made of unknown non-baryonic particles (for a brief review on dark matter see for instance: Einasto, 2010). A full list of the proposed dark matter particles would be longer than this letter; let us mention only weekly interacting massive particles and axions. In spite of the significant efforts dark particles have never been detected. Let us note that in order to fit observational data for a galaxy, the radial mass density of dark matter in a halo should be nearly constant

$$\rho_r = \frac{dM_{dm}}{dr} \approx const. \tag{1}$$

The best developed alternative to particle dark matter is the Modified Newtonian Dynamics (MOND). It states that there is no dark matter and we witness a violation of the fundamental law of gravity (see brief review of Milgrom, 2010).

In a recent series of papers (Blanchet 2007a, 2007b; Blabchet and Tiec 2008, 2009) it was shown that, in spite of the fact that MOND phenomenology rejects the existence of dark matter, it can be considered as consequence of a particular form of dark matter. The key hypothesis is that dark matter is a dipolar fluid composed from gravitational dipoles (in analogy with electric dipole, a gravitational dipole is defined as a system composed of two particles, one with positive and one with negative gravitational charge). Hence, Blanchet and Tiec have introduced dipolar fluid as a new candidate for non-baryonic dark matter; the galaxy rotation curves can be considered as result of the gravitational polarization of the dipolar fluid by the gravitational field of baryonic matter.

While the work of Blanchet and Tiec has attracted a significant attention, a very different idea concerning the gravitational polarization (Hajdukovic, 2007, 2008) passed in silence. The key hypothesis advocated by Hajdukovic is the gravitational repulsion between matter and antimatter, i.e. particles and antiparticles have gravitational charge of opposite sign. Consequently the virtual particle-antiparticle pairs in the quantum vacuum should be considered as gravitational dipoles. Thus, the quantum vacuum may be considered as a dipolar fluid, what is much simpler and more elegant than the dipolar fluid composed from the unknown non-baryonic matter. As we would argue in this letter, in the framework of this approach, dark matter does not exist but is an illusion created by the polarization of the quantum vacuum by the gravitational field of the baryonic matter. Hence, for the first time, the quantum vacuum fluctuations, well established in quantum field theory but mainly neglected in astrophysics and cosmology, are related to the problem of dark matter.

The existing experimental evidence and our assumption of gravitational repulsion between matter and antimatter may be summarized as:

$$m_i = m_g \; ; \; m_i = \overline{m}_i \; ; \; m_g + \overline{m}_g = 0 \tag{2}$$

Here, as usually, a symbol with a bar denotes antiparticles; while indices *i* and *g* refer to inertial and gravitational mass (gravitational charge). The first two relations in (2) are experimental evidence (Will, 1993; Gabrielse, 1999), while the third one is our assumption which dramatically differs from general conviction that $m_g - \overline{m}_g = 0$. Our hypothesis was very recently supported by a striking result (Villata, 2011) that "antigravity appears as a prediction of general relativity when CPT is applied".

According to hypothesis $m_g + \overline{m}_g = 0$, a virtual pair may be considered as gravitational dipole with the gravitational dipole moment

$$\vec{p} = m\vec{d}; \; |\vec{p}| < \frac{\hbar}{c} \tag{3}$$

Here, by definition, the vector $\vec{d}$ is directed from the antiparticle to the particle, and presents the distance between them. Consequently, a gravitational polarization density $\vec{P}_g$ (i.e. the gravitational dipole moment per unit volume) may be attributed to the quantum vacuum. The inequality in (3) follows from the fact that distance between virtual particle and antiparticle must be smaller than the reduced Compton wavelength $\mathchar'26\mkern-10mu\lambda_m = \hbar/mc$ (for larger separations a virtual pair becomes real). Hence, $|\vec{p}|$ should be a fraction of $\hbar/c$.

In quantum field theory, a virtual particle-antiparticle pair (i.e. a gravitational dipole) occupies the volume $\lambda_m^3$, where $\lambda_m$ is the (non-reduced) Compton wavelength. Hence, the number density of the virtual gravitational dipoles has constant value

$$N_0 \approx \frac{1}{\lambda_m^3} \tag{4}$$

In order to grasp the key difference between the polarization by an electric field and the eventual polarization by a gravitational field, let's remember that, as a consequence of polarization, the strength of an electric field is reduced in a dielectric. For instance, when a slab of dielectric is inserted into a parallel plate capacitor, the electric field between plates is reduced. The reduction is due to the fact that the electric charges of opposite sign attract each other. If, instead of attraction, there was repulsion between charges of opposite sign, the electric field inside a dielectric would be augmented. But, according to our hypothesis, there is such repulsion between gravitational charges of different sign. Consequently, outside of a region in which a certain baryonic mass $M_b$ is confined, the eventual effect of polarization should be a gravitational field stronger than predicted by the Newton's law (but without violation of the Newton law in the same way as electric polarization is not violation of the Coulomb law). In more technical words we have the case of anti-screening by virtual particle-antiparticle pairs. The most important question is if the gravitational polarization of the vacuum can produce the same effect as the presumed existence of dark matter.

If the quantum vacuum "contains" the virtual gravitational dipoles, a massive body with mass $M_b$ (a star, a black hole…), but also multi-body systems as galaxies should produce vacuum polarization, characterized with a gravitational polarization density $\vec{P}_g$.

As well known, in a dielectric medium the spatial variation of the electric polarization generates a charge density $\rho_b = -\nabla \cdot \vec{P}$, known as the bound charge density. In an analogous way, the gravitational polarization of the quantum vacuum should result in a gravitational bound charge density of the vacuum

$$\rho_v = -\nabla \cdot \vec{P}_g \tag{5}$$

If we assume the spherical symmetry, the equation (5) may be reduced to

$$\rho_v(r) = \frac{1}{r^2}\frac{d}{dr}\left(r^2 P_g(r)\right); \quad P_g(r) \equiv |\vec{P}_g(r)| \geq 0 \tag{6}$$

or, as we are interested in the radial gravitational charge density (as in the equation (1))

$$\rho_r = 4\pi \frac{d}{dr}\left(r^2 P(r)\right) \tag{7}$$

In principle the space around a spherical body can be divided in 3 regions which we define with two critical radiuses denoted by $R_0$ and $R_h$, ($R_0 < R_h$); the values of $R_0$ and $R_h$ would be estimated later.

In the region ($r < R_0$) the gravitational field is sufficiently strong to align all dipoles along the field and consequently $P_g(r)$ has a constant value (in fact a maximum value) which according to equations (3) and (4) may be written as

$$P_g(r) \equiv P_{g\,max} = \frac{A}{\lambda_m^3}\frac{\hbar}{c} \tag{8}$$

where A should be a dimensionless constant of order of unity. The equations (7) and (8) lead to the radal gravitational charge density proportional to $r$, producing a constant radial acceleration towards the body, which may be related to the Pioneer anomaly (Hajdukovic 2010a).

For $r > R_h$ the gravitational field is so weak that dipoles are randomly oriented and hence $P_g(r)$ is zero (we may also allow non-zero values, for instance $P_g(r)$ decreasing as $1/r^2$ or faster but it is not important for the present study).

Inside the spherical shell (with the inner radius $R_0$ and the outer radius $R_h$) the external gravitational field is not sufficiently strong to align all dipoles, but also not so weak to allow random orientation; hence the polarization density $P_g(r)$ should decrease with distance. Only in the region of this spherical shell, we may attempt to describe phenomena by the gravitational polarization instead of particle dark matter.

With

$$P_g(r) \sim P_{g\,max}\frac{R_0}{r} \tag{9}$$

equation (7) leads to

$$\rho_r \sim 4\pi P_{g\,max} R_0 = 4\pi A \frac{\hbar}{c}\frac{R_0}{\lambda_m^3} \tag{10}$$

Now, it is necessary to estimate $\lambda_m$ and $R_0$ in the above relation.

Recently, two independent approaches (Urban and Zhitnitsky 2009, 2010 and Hajdukovic 2010b, 2010c, 2010d) have supported the point of view that only QCD (quantum chromodynamics) vacuum is significant for gravitation (Let us note that speculations concerning the gravitational properties of the quantum vacuum have their roots in the work of Zeldovich 1967). Roughly speaking the QCD vacuum is a gas of virtual pions (quark-antiquark pairs);

consequently $\lambda_m$ in equation (10) should be identified with the Compton wavelength of pion ($\lambda_m = \lambda_\pi$).

In order to estimate $R_0$, let us note that according to equation (4), $\lambda_\pi$ can be interpreted as the mean distance between two dipoles which are the first neighbors. The gravitational acceleration produced by a pion at the distance of its own Compton wavelength is:

$$a_o = \frac{Gm_\pi}{\lambda_\pi^2} = \left(\frac{c}{h}\right)^2 Gm_\pi^3 \approx 2\times 10^{-10} m/s^2 \qquad (11)$$

what is intriguingly close to the value of the fundamental acceleration conjectured in the MOND phenomenology.

In the region $r < R_0$, the acceleration (11) should be dominated by the acceleration produced by the baryonic mass $M_b$. Hence, $R_0$ may be estimated from the condition of equality between $a_0$ and the Newtonian acceleration $GM_b/R_0^2$

$$R_0 = \lambda_\pi \sqrt{\frac{M_b}{m_\pi}} \qquad (12)$$

For a mass $M_b \approx 4\times 10^{41} kg$ (corresponding to the mass of our galaxy) the numerical value is $R_0 \approx 3.55\times 10^{20} m \approx 11.5 kpc$.

Including these estimations, relation (10) may be written as

$$\boxed{\rho_r = \frac{B}{\lambda_\pi}\sqrt{m_\pi M_b}} \qquad (13)$$

where $B$ is a dimensionless constant of order of unity.

The relation (13) is an intriguingly simple rule: find the geometrical mean of the mass of pion and baryonic mass of a galaxy and divide it with the Compton wavelength of pion, what you get is the order of the radial dark matter density. An additional feature is that equation (13) leads to the Tully-Fisher empirical relation between the asymptotic flat velocity and the luminosity of spirals (Tully and Fisher, 1977). In fact, for large radii, the equation (13) together with the well known result for rotational velocity at a circular orbit, $V_{rot}(r) = \sqrt{GM(r)/r}$, leads to

$$V_{rot}^4 = G^2 \rho_r^2 = B^2 G^2 \frac{m_\pi M_b}{\lambda_\pi^2} \qquad (14)$$

what is (assuming proportionality between the luminosity and baryonic mass ) Tully-Fisher relation $L \sim V_{rot}^4$. Let us note that in a recent paper (McGaugh, 2011) it was argued that the Tully-Fisher relation is universally valid for all types of galaxies.

Introducing the ratio

$$\frac{M_{dm}}{M_b} = \frac{\Omega_{dm}}{\Omega_b} \qquad (15)$$

where dimensionless parameters $\Omega_{dm}$ and $\Omega_b$ denote "dark matter" and baryonic matter density, the equation (13) leads to the following estimation of the size of "dark matter" halo

$$R_h = \frac{1}{B}\frac{\Omega_{dm}}{\Omega_b} R_0 + R_0 = \frac{1}{B}\frac{\Omega_{dm}}{\Omega_b} \lambda_\pi \sqrt{\frac{M_b}{m_\pi}} + R_0 \qquad (16)$$

As observations suggest, the ratio $\Omega_{dm}/\Omega_b$ is a little bit smaller than $5$, while $R_h$ is presumably more than $20$ times larger than $R_0$; hence according to (16), $B$ must have a value close to $\Omega_b/\Omega_{dm}$. In principle, observational data may serve to determine the appropriate value of $B$, but the trouble is that they are not very accurate; for instance the halo's viral mass of our Galaxy has not be constrained to better than a factor of 2-3. Taking again $M_b \approx 4 \times 10^{41} kg$ and $B = \Omega_b/\Omega_{dm} \approx 0.212$, formula (16) gives $R_h \approx 266 kpc$ what is a surprisingly good result for the halo size of our Galaxy (let us remember that we have used a toy model with spherical symmetry not taking into account real distribution of baryonic matter in galaxy).

Let us give one more numerical illustration concerning our galaxy. Xue et al. (2008) have found that the mass enclosed within $60 kpc$ is $(8 \pm 1.4) \times 10^{41} kg$, while our toy model estimate is $7.7 \times 10^{41} kg$.

In conclusion, we have revealed the first indications that what we call dark matter may be consequence of the gravitational repulsion between matter and antimatter and the corresponding gravitational polarization of the quantum vacuum by the existing baryonic matter. Of course, this is not a claim, just possibility. A lot of work would be needed before such a claim would be eventually possible. Our work is in progress to see if the formalism developed by Blanchet and Tiec can be applied in our case and produce accurate results in the framework of General Relativity.

Let us end by pointing that the rotational curves of galaxies are not the only phenomenon which is currently explained by Dark Matter. For instance, CMB data are apparently in favor of the presence of dark matter as a key for understanding of density fluctuations and the structure formation in the Universe (see review of Einasto, 2010). While our Letter gives indices that the gravitational vacuum polarization could be an alternative to dark matter in the explanation of the galactic rotational curves, a tremendous work would be needed, to reveal if the other phenomena could be alternatively explained by the vacuum polarization.

# Appendix A
# Quantum vacuum and dark matter

Dragan Slavkov Hajdukovic[1]
PH Division CERN
CH-1211 Geneva 23
dragan.hajdukovic@cern.ch
[1]On leave from Cetinje, Montenegro

**Abstract**
Recently, the gravitational polarization of the quantum vacuum was proposed as alternative to the dark matter paradigm. In the present paper we consider four benchmark measurements: the universality of the central surface density of galaxy dark matter haloes, the cored dark matter haloes in dwarf spheroidal galaxies, the non-existence of dark disks in spiral galaxies and distribution of dark matter after collision of clusters of galaxies (the Bullet cluster is a famous example). Only some of these phenomena (but not all of them) can (in principle) be explained by the dark matter and the theories of modified gravity. However, we argue that the framework of the gravitational polarization of the quantum vacuum allows the understanding of the *totality* of these phenomena.

1. Introduction

Contemporary physics has two cornerstones: General Relativity and the Standard Model of Particle Physics. General Relativity is our best theory of gravitation. The Standard Model is a collection of Quantum Field Theories; according to the Standard Model, everything in the Universe is made from six quarks and six leptons (and their antiparticles) which interact through exchange of gauge bosons (photon for electromagnetic interactions, $W^{\pm}$ and $Z^0$ for weak interactions and eight gluons for strong interactions).

The problem is that our best physics is apparently insufficient to explain a series of major phenomena discovered in Astrophysics and Cosmology. One of the unexplained phenomena is that the gravitational field in the Universe is much stronger than it should be according to our theory of gravity and the existing amount of the baryonic matter (i.e. the matter composed from the Standard Model particles). This phenomenon is considered as a strong hint that at least one of cornerstones (General Relativity and Standard Model) must be significantly modified. Both approaches (modification of the fundamental law of gravity and the assumption that in addition to quarks and leptons there are still unknown fundamental particles named dark particles) have been studied by thousands of scientists, but a solution is still not at hand.

Recently (Hajdukovic, 2011; but see also the first appearance of the idea in Hajdukovic, 2007 and Hajdukovic, 2008)) a third way, without invoking dark matter and without invoking the modification of the fundamental law of gravity, has been proposed. In simple words, according to the Quantum Field Theory, all baryonic matter in the Universe is immersed in quantum vacuum; popularly speaking a "sea" of short living virtual particle-antiparticle pairs (like electron-positron pairs with the lifetime of about $10^{-22}\,s$, or neutrino-antineutrino pairs with a lifetime of about $10^{-15}\,s$ which is a record lifetime in the quantum vacuum). It is difficult to believe that quantum vacuum does not interact gravitationally with the baryonic matter immersed in it. In spite of it, the quantum vacuum is ignored in astrophysics and cosmology; not because we are not aware of its importance but because no one has any idea what the gravitational properties of the quantum vacuum are. In absence of any knowledge, as a starting point, we have conjectured that particles and antiparticles have the gravitational charge of opposite sign. An immediate consequence is the existence of the gravitational dipoles; a virtual pair is a gravitational dipole (in the same way as a virtual electron-positron pair is an electric dipole), that allows the gravitational polarization of the quantum vacuum. The initial study (Hajdukovic, 2011) has revealed the surprising possibility that the gravitational polarization of the quantum vacuum can produce phenomena usually attributed to dark matter. In the present paper we focus on four benchmark phenomena established by observations: (a) the universality of the central surface density of galaxy dark matter haloes (Donato et al. 2009), (b) the cored dark matter haloes in dwarf spheroidal galaxies (Walker and Penarrubia, 2011), (c) the non-existence of dark disks in spiral galaxies (Moni Bidin et al. 2010) and (d) the distribution of dark matter after collisions of clusters of galaxies (the Bullet cluster (Clove et al. 2006) being a famous example). In section 2 we give a brief review of these four phenomena and point to the known fact that only some of them (but not all of them) can in principle be explained by the dark matter and the modified theories of gravity. In section 3 we consider the same phenomena in the framework of the gravitational polarization of the quantum vacuum and argue that it is the framework in which the totality of these phenomena can be understood. Section 4 is devoted to discussion.

## 2. Four important measurements

Let us give a brief review of four observed phenomena which have become benchmark for different theories. Both, the cold dark matter model and MOND fail to explain the totality of these phenomena. The dark matter theory has more problems at small scales, while modified gravity (we take MOND as leading example) has significant problems at large scales.

*(a) Central surface density*

There is strong evidence (Donato et al. 2009) that the central surface density $\mu_{0D} \equiv r_0 \rho_0$ of galaxy dark matter haloes (where $r_0$ and $\rho_0$ are the halo core radius and central density) is nearly constant and independent of galaxy luminosity. The measured value (Donato et al. 2009) is about 140 solar masses per square parsec

$$\mu_{0D} \equiv r_0 \rho_0 = 140_{-30}^{-80} \frac{M_{Sun}}{pc^2} = 0.29 \frac{kg}{m^2} \quad (1)$$

The universality of the dark matter surface density at the core radius is a mystery for the particle dark matter but can be explained within the MOND phenomenology (Milgrom, 2009). As we will see, the gravitational polarization of the quantum vacuum obviously leads to a relation producing the numerical result (1).

*(b) Dwarf spheroidal galaxies*

Dwarf spheroidal galaxies, with a typical diameter of about 1000 light years, are the smallest galaxies observed in the Universe. For a number of reasons they are considered as an important "laboratory" for the study of dark matter distribution at the centres of galaxies. Recently, Walker and Penarrubia (2011) have accomplished the first direct measurements that reveal how densely dark matter is packed toward the centres of two nearby dwarf galaxies (Fornax and Sculptor) that orbit the Milky Way as satellites.

The measured slope

$$\Gamma \equiv \frac{\Delta \log M}{\Delta \log r} \quad (2)$$

is $\Gamma \approx 2.61$ and $\Gamma \approx 2.95$ respectively for Fornax and Sculptor galaxy. The values of $\Gamma$ in the range $2 < \Gamma < 3$, are consistent with cored dark matter halos of an approximately constant density over the central few hundred parsecs, what contradicts the cusp distribution ($\Gamma < 2$) predicted by the current cold dark matter theory.

Hence, Walker and Penarrubia have provided the first direct evidence that the cold dark matter paradigm cannot account for the phenomenology of dark matter at small scales.

*(c) Dark disks*

Everyone knows that our Galaxy is immersed in a halo of dark matter (a real one if we trust the cold dark matter theory or a phantom halo according to theories of modified gravity like MOND). It is less known that in addition to the halo, our galaxy should have a dark matter disk, which is thicker than the visible galactic disk. The presence of a real dark disk is a natural expectation of the cold dark matter model (Read et al. 2008) while the presence of a phantom disk (Milgrom, 2001) is a prediction of MOND theory. The observations suggest (Moni Bidin et al. 2010) that at this point both theories are wrong; apparently, dark matter disk does not exist. As we will show in Section 3, the non-existence of dark matter disk is a natural consequence of the gravitational polarization of the quantum vacuum.

*(d) The Bullet cluster*

The observations of the Bullet cluster show the distribution of the baryonic and dark matter after collision of two clusters of galaxies.

During the collision, the galaxies within the two clusters passed by each other without interactions (because of the large distances between them), while the interacting clouds of X-ray emitting plasma have been slowed by ram pressure. Hence, two clouds of plasma are now located between the two separated clusters. The key point is that the distribution of dark matter (determined by the gravitational lensing) is centred on clusters, while the dominant part of baryonic matter is in clouds of plasma. Such a common "destiny" of dark matter and stellar components of clusters can't be explained by modified gravity where dark matter should be centred on the dominant part of the baryonic matter (i.e. on clouds of plasma). However, in the framework of the cold dark matter theory, dark matter is collisionless and it is natural that it behaves in the same way as the collisionless part of the baryonic matter.

3. Gravitational polarization of the quantum vacuum

Let us assume that particles and antiparticles have the gravitational charge of the opposite sign. Consequently, a virtual particle-antiparticle pair may be considered as a gravitational dipole with the gravitational dipole moment

$$\vec{p} = m\vec{d}; \; |\vec{p}| < \frac{\hbar}{c} \qquad (3)$$

Here, by definition, the vector $\vec{d}$ is directed from the antiparticle to the particle, and presents the distance between them. The inequality in (3) follows from the fact that the distance between virtual particle and antiparticle must be smaller than the reduced Compton wavelength $\lambdabar_m = \hbar/mc$ (for larger separations a virtual pair becomes real). Hence, $|\vec{p}|$ should be a fraction of $\hbar/c$.

If the quantum vacuum "contains" the virtual gravitational dipoles, the gravitational field of a body immersed in the quantum vacuum, should produce vacuum polarization, characterized with a gravitational polarization density $\vec{P}_g$ (i.e. the gravitational dipole moment per unit volume).

In the quantum field theory, a virtual particle-antiparticle pair (i.e. a gravitational dipole) occupies the volume $\lambda_m^3$, where $\lambda_m$ is the (non-reduced) Compton wavelength. As argued in previous papers (Hajdukovic 2010, Hajdukovic 2011) the pions (as the simplest quark-antiquark pairs) dominate the quantum vacuum and $\lambda_m$ should be identified with the Compton wavelength $\lambda_\pi$ of a pion. Hence, the number density of the virtual gravitational dipoles has a constant value

$$N_0 \propto \frac{1}{\lambda_\pi^3} \qquad (4)$$

According to equations (3) and (4), if all dipoles are aligned in the same direction, the gravitational polarization density $\vec{P}_g$ has the maximal magnitude

$$\left|\vec{P}_g\right| \equiv P_{g\,max} = \frac{A}{\lambda_\pi^3}\frac{\hbar}{c} \qquad (5)$$

where $A < 1$, should be a dimensionless constant of order of unity. This may happen only in a sufficiently strong gravitational field with magnitude $g$, larger than a critical value $g_{cr}$.

The critical field $g_{cr}$ should have the same order of magnitude (Hajdukovic, 2011) as the gravitational acceleration produced by a pion at the distance of its own Compton wavelength

$$g_{cr} = B\frac{Gm_\pi}{\lambda_\pi^2} = 2.1B \times 10^{-10}\,m/s^2 \qquad (6)$$

where $B$ is a dimensionless constant of order of unity. The numerical value of $g_{cr}$ is surprisingly close to the fundamental acceleration $a_0$ conjectured by MOND; in fact $g_{cr} = a_0$ implies $B \approx 0.58 \approx 1/\sqrt{3}$ and we will adopt this value for $B$ in numerical calculations. The fact that a critical acceleration appears in our theory is only a superficial similarity with MOND; in our approach there is no modification of the fundamental law of gravity for $g < g_{cr}$.

The equations (5) and (6), together with the proportionality

$$P_{g\,max} = \frac{1}{4\pi G}g_{cr} \qquad (7)$$

lead to $2A = B$, i.e.

$$A \approx 0.29 \approx \frac{1}{2\sqrt{3}};\quad B \approx 0.58 \approx \frac{1}{\sqrt{3}} \qquad (8)$$

Let us note that $1/4\pi G$ plays the role of the gravitational vacuum permittivity, analogous to the vacuum permittivity $\varepsilon_0$ in electrodynamics).

As previously suggested (Hajdukovic, 2011), dark matter density may be interpreted as the density of the gravitational polarization charges.

$$\rho_{dm} = -\nabla \cdot \vec{P}_g \qquad (9)$$

If we assume the spherical symmetry, (9) may be reduced to

$$\rho_{dm}(r) = \frac{1}{r^2}\frac{d}{dr}\left(r^2 P_g(r)\right) \qquad (10)$$

with $P_g(r) \equiv \left|\vec{P}_g(r)\right|$.

Let us note that from the purely mathematical point of view there are three interesting possibilities: $P_g(r)$ is directly proportional to $r$, $P_g(r) = const$ and $P_g(r)$ is inversely proportional to $r$. In these particular cases, the equation (10) leads respectively to the constant volume density, constant surface density and constant radial density of dark matter, i.e.

$$P_g(r) \propto r \Rightarrow \frac{dM_{dm}}{dV} = C_1 \qquad (11)$$

$$P_g(r) = const \Rightarrow \frac{dM_{dm}}{dS} = C_2 \qquad (12)$$

$$P_g(r) \propto \frac{1}{r} \Rightarrow \frac{dM_{dm}}{dr} = C_3 \qquad (13)$$

where $C_1$, $C_2$ and $C_3$ are some constants. We will see that all these mathematical possibilities approximate the real physical situations. Let us note that we continue to use the words dark matter, while it is not more the dark matter of unknown nature, but the effect of the rearrangement of the virtual gravitational charges in the quantum vacuum.

In fact, the case (13) was already studied (Hajdukovic, 2011), leading to the main result:

$$\frac{dM_{dm}(r)}{dr} = \frac{B}{\lambda_\pi}\sqrt{m_\pi M_b} \qquad (14)$$

describing a dark matter halo outside of a spherically symmetric distribution of the baryonic mass $M_b$; a result that mimics well the observed galactic dark matter halo at relatively large distances from the center of the galaxy. Hence, in the present paper we will focus on the cases (11) and (12).

As an example of the baryonic distribution without spherical symmetry, let us consider a planar-like distribution (like for instance a thin galactic disk). From the mathematical point of view, the simplest case is an infinite plane with a constant baryonic surface mass density $\sigma_b$ (this is the gravitational version of an infinite plane with constant electric charge density, what is an exercise known to every student of physics). The gravitational field $\vec{g}$ produced by the plane is perpendicular to the plane, oriented towards the plane and has a constant magnitude which can be determined by a trivial application of the Gauss's flux theorem

$$|\vec{g}| = 2\pi G \sigma_b \qquad (15)$$

In a constant gravitational field $\vec{g}$ the gravitational polarization density $\vec{P}_g$ should be a constant vector and its divergence (i.e. the right-hand side of the equation (9)) is zero. Hence, while the vacuum around the considered plane is polarized, dark matter density is zero. Consequently, close to a large plane or between two large planes, there is no significant gravitational field caused by the gravitational polarization of the quantum vacuum. By the way, it leads to the conclusion that the baryonic galactic disk of our galaxy can't be accompanied by a thicker dark matter galactic disk, what contradicts the common prediction of the cold dark matter theory (Read et al. 2008) and MOND (Milgrom, 2009) . Recent studies (Bidin et al., 2010) show that there is no evidence for a dark matter disk within 4 kpc from the galactic plane, which apparently confirm our prediction.

The above considerations suggest that we may live in a Universe with a variable quotient of the baryonic and dark matter. To see it, let us imagine, that a spherical distribution of baryonic matter is somehow "deformed" to a planar-like distribution. In these two cases, a distinct observer would measure the same quantities of baryonic matter, but different quantities of dark matter!

*3.1 Gravitational field stronger than the critical value*

Let us turn back to the case of spherical symmetry. In general, there are two regions outside a distribution of the baryonic matter; the region with $g \geq g_{cr}$ and the region with $g < g_{cr}$.

The region with $g \geq g_{cr}$ is the easiest for the study; we have the estimate (5) for the maximal magnitude of the gravitational polarization density and we can use it in the equation (10), without need for a detailed understanding of the quantum vacuum, what is the major problem in the case $g < g_{cr}$. It is evident that the mathematical case (12) corresponds to the physical case when the gravitational field is sufficiently strong to produce saturation. From (5) and (10) it is easy to obtain the relation

$$r\rho_{dm}(r) = 2P_{g\,\max} = \frac{2A}{\lambda_\pi^3}\frac{\hbar}{c} \equiv \frac{A}{\pi}\frac{m_\pi}{\lambda_\pi^2} \qquad (16)$$

which explains the observed universality (1) of the central surface density and gives (using the value of $A$ determined in (8)) a numerical value in the excellent agreement with the measurements. Alternatively we may consider the measurement (1) as the experimental determination of the constant $A$ in equations (5) and (16).

Let us forget for the moment how we have obtained the result (16). Even if considered in isolation, as an ad hoc formula, it is astonishing that a universal quantity as (1) can be expressed through universal constants and mass of a quark-antiquark pair (what is roughly a pion).

According to (16) the mass of dark matter enclosed inside a sphere with radius $r$ is

$$M_{dm}(r) = Bm_\pi \left(\frac{r}{\lambda_\pi}\right)^2 \qquad (17)$$

while the acceleration produced by the dark matter has a constant value equal to the critical acceleration.

$$g_{dm}(r) = \frac{GM_{dm}(r)}{r^2} \equiv g_{cr} \qquad (18)$$

So, in the region $g > g_{cr}$, the total acceleration at distance $r$ is the sum of the acceleration $g_b(r)$ caused by the baryonic matter (and described by the Newton law), and a very small, constant acceleration (18) caused by the dark matter and oriented towards the center of the spherical symmetry. In the region $g < g_{cr}$, $g_{dm}(r)$ is not more a constant, but depends on $r$ what can be wrongly interpreted as a modification of the Newton law (a mistake included as cornerstone of the MOND phenomenology).

By the way, the additional constant sunward acceleration (18) should exist in the Solar system and affect the orbital motions of the Solar system's bodies, but in order to detect it, we must know orbits with higher accuracy (which may be not so far into the future; Page et al. 2009)

*3.2. Gravitational field weaker than the critical value*

For bodies like a star or our planet, the gravitational field becomes stronger than the critical one in less than one meter from the center of the body. Hence, the gravitational field around a star has an inner region with $g > g_{cr}$, and an outer region with $g < g_{cr}$. The region $g > g_{cr}$ should be called the region of saturation because the polarization density has a maximal magnitude. The same should be true for a Galaxy with a supermassive black hole in the center. For instance, the supermassive black hole in the centre of the Milky Way assures condition $g > g_{cr}$ at distances of more than 100 light years (without counting other baryonic matter in the central region).

The other possibility is the existence of a large central region with $g < g_{cr}$. It is possible if there is a sufficiently low baryonic mass density in the central part of a galaxy.

Let us consider a sphere filled with the baryonic matter of the volume density $\rho_b(r)$ which depends only on the distance $r$ from the centre. The gravitational acceleration produced by the baryonic matter is

$$g_b(r) = \frac{4\pi G}{r^2} \int_0^r r^2 \rho_b(r) dr \qquad (19)$$

It is obvious that an analogous relation exists for the acceleration $g_{dm}(r)$ produced by the dark matter. In the particular case of an approximately constant baryonic volume density $\rho_b(r) \equiv \rho_b$, the equation (19) leads to the direct proportionality between acceleration $g_b(r)$ and the radial distance $r$, i.e.

$$g_b(r) = \frac{4\pi G \rho_b}{3} r \qquad (20)$$

According to (20), the assumption of the direct proportionality between $P_g(r)$ and $g(r)$ means that $P_g(r)$ is also proportional to $r$, what corresponds to the mathematical case (11), describing a cored dark matter halo.

However, at this point the problem is that we do not know the properties of the quantum vacuum and in particular we do not know if for $g < g_{cr}$, the magnitude of polarization density grows with the acceleration in a linear or non-linear manner.

To be more general, let us assume a non-linear growth of the polarization density

$$P_g(r) = Kr^x \qquad (21)$$

where $K$ is a constant and $x \leq 1$ a positive number. Using this form for $P_g(r)$ in the basic equation (10) and after that using the obtained result to calculate the slope (2), leads to $\Gamma = 2 + x$, i.e. $2 < \Gamma \leq 3$, as observed for dwarf spheroidal galaxies (Walker and Panarrubia, 2011)

*3.3 The Bullet cluster*

Because of the mathematical complexity, the numerical simulations are inevitable and crucial in our present day studies of dark matter. A simulation of the Bullet cluster (and some other problems) in the framework of the gravitational polarization of the quantum vacuum is an urgent task. However it is easy to see that the observed separation of dark matter and the dominant part of the baryonic matter is not a surprise.

The key question is why there is no significant presence of dark matter between the clouds of the X-ray emitting plasma. First, while the three dimensional form of clouds is not known, during the collision the clouds were not only slowed but flattened as well. And, as we have argued above, around a flattened distribution of the baryonic matter, the additional field caused by the gravitational polarization is not significant. The second important factor is that the distance between clouds is relatively small. When two baryonic masses are close enough, they compete to orient the same dipoles in different directions, what changes the gravitational polarization density and its divergence. Hence, while without the appropriate simulations a detailed picture is impossible, the absence of dark matter in the region of clouds has nothing unusual.

4. Discussion

The initial paper (Hajdukovic, 2011) has revealed an intriguingly simple rule: find the geometrical mean of the mass of a pion and the baryonic mass of a galaxy and divide it with the Compton wavelength of the pion; what you get is very close to the observed radial dark matter density in a galaxy (see the equation (14)). It was the first indication that what we call dark matter may be the result of the gravitational polarization of the quantum vacuum.

In the present paper we have revealed the additional indications; the most striking one is the result of equation (16), a universal property of galaxies (1) can be expressed through the universal constants and mass of pion what is simply astonishing. There is one point here which deserves particular attention. The Planck constant $\hbar$, so crucial in quantum theory, but absent from our theory of gravitation, appears in both equations (14) and (16) concerned with the large scale gravitational phenomena. All this suggests that the gravitational polarization of the quantum vacuum may be a serious alternative to the dark matter paradigm.

Let us clarify that our theory is not a support to MOND. Yes, there is a critical gravitational field; in a field stronger than the critical one there is saturation (i.e. the maximal gravitational polarization density), but there is no violation of the fundamental law of gravity. The fact that MOND correctly guessed the existence of a critical field is the reason for its partial success, but (in our opinion which may be wrong) the success is limited because of the misunderstanding of the physical origin of this critical field.

Let us end with one intriguing question. Are the result (16) and its consequence (17) valid at the scale of the whole Universe? The answer may be yes. Let us use in the equation (17) the radius of the observable Universe, which is estimated to be about 14 billion parsecs i.e. $\approx 4.3 \times 10^{26} m$. According to (17) the corresponding dark matter in the Universe is about $3.4 \times 10^{53} kg$ or $1.7 \times 10^{23}$ solar masses. If our estimate of the current ratio of the baryonic and dark matter in the Universe is correct, the baryonic mass of the visible universe should be $3 \times 10^{22}$ solar masses. Everything looks as if equation (17) is valid for the Universe as a whole. But if so, the ratio of the dark matter and the baryonic matter in the universe should grow with time.